\documentclass[10pt,aps,physrev,twocolumn,groupedaddress]{revtex4-2}
\usepackage{graphicx}
\usepackage{amsmath}
\usepackage[caption=false]{subfig}

\begin{document}

\title{Scalar curvature density as a new invariant in thermodynamic geometry: metric dependence and critical exponents}

\author{José Torres-Arenas}
\affiliation{División de Ciencias e Ingenierías, Universidad de Guanajuato}
\affiliation{Departamento de Física Aplicada, Universidade de Vigo}

\author{Jaime Jaramillo-Gutiérrez}
\email{jaime.jaramillo@ugto.mx} 
\affiliation{División de Ciencias e Ingenierías, Universidad de Guanajuato}

\author{Juan Becerra-Zamudio}
\affiliation{División de Ciencias e Ingenierías, Universidad de Guanajuato}

\date{\today}

\begin{abstract}

We compare two Ruppeiner metrics constructed under fixed volume and fixed particle number conditions ($g_{_V}$ and $g_{_V}$) by analyzing the scalar curvature $R$ and introducing the scalar curvature density $\mathcal{R} = \sqrt{|g|}\,R$ as a complementary geometric invariant. Three fluid models of increasing
physical realism are considered: van der Waals, Lennard-Jones, and argon described by a multiparameter equation of state that correctly reproduces non-mean-field critical behavior consistent with the Ising universality class.
We find that $R$ and $\mathcal{R}$ exhibit distinct critical scaling: $R \sim t^{-d\nu}$ is governed by the correlation length exponent, whereas $\mathcal{R} \sim t^{-(1+\beta)}$ scales solely with the order parameter
exponent $\beta$, a result that follows analytically from hyperscaling and the Rushbrooke relation independently of the universality class. The $N$-metric consistently outperforms the $V$-metric in reconstructing the vapor-liquid
coexistence curve away from criticality, and the four Widom lines defined by the minima of $R_{_V}$, $R_{_N}$, $\mathcal{R}_{_V}$, and $\mathcal{R}_{_N}$ display a characteristic fourfold structure reproduced across all three models. The loci where both representations yield identical geometric descriptions define two new
objects: the Curvature Equality Curve (CEC, $R_{_V} = R_{_N}$) and the Curvature-Density Equality Curve (CDEC, $\mathcal{R}_{_V} = \mathcal{R}_{_N}$), with the CDEC enclosing a substantially larger region of the phase diagram in all cases. These results establish $\mathcal{R}$ as a meaningful complement to $R$ in thermodynamic geometry and highlight the nontrivial role of metric choice in the description of phase transitions and supercritical behavior.
\end{abstract}

\maketitle

\section{Introduction}
Thermodynamic geometry has emerged as a powerful framework to characterize interacting many-body systems by endowing the space of equilibrium states with a Riemannian structure. In this approach, originally introduced by Ruppeiner \cite{Ruppeiner1979,Ruppeiner1995}, the metric is defined as the Hessian of the entropy (or equivalently of a thermodynamic potential), and its associated scalar curvature provides information about the underlying microscopic interactions. In particular, the thermodynamic curvature has been widely interpreted as a measure of the correlation volume \cite{Ruppeiner2010,Ruppeiner2012}, establishing a bridge between macroscopic thermodynamics and microscopic structure.\\

A subtle but important aspect of this construction is that the resulting geometry is not unique. The definition of the metric depends on the choice of thermodynamic potential and, crucially, on the set of independent variables used to describe the system \cite{Ruppeiner1979,Ruppeiner1995,Vega2022,Rodrigo2025,Quevedo2007,Quevedo2008,Mahmoudi2023}. In practice, this implies that different ensemble constraints induce different metrics on the thermodynamic state space. For instance, fixing the volume or fixing the number of particles leads to distinct geometric structures, commonly denoted as $g_{_V}$ and $g_{_N}$ \cite{Vega2022,Rodrigo2025}. Although both describe the same physical system, their associated scalar curvatures generally differ, reflecting the fact that they encode different fluctuation structures.\\

Despite this intrinsic ensemble dependence, most studies adopt a single metric and do not systematically explore how the choice of ensemble affects the resulting geometric description. This raises several natural questions: to what extent are geometric features such as curvature extrema, coexistence conditions, or supercritical structures dependent on the chosen metric? Are there regions of the phase diagram where different metrics become effectively equivalent?
In this work, we address these questions through a systematic comparison of the metrics $g_{_V}$ and $g_{_N}$ in simple fluids. As a starting point, we consider the van der Waals fluid, which provides a simple and analytically tractable framework to expose ensemble-dependent geometric features. While its thermodynamic geometry has been widely studied, previous analyses have largely focused on $g_{_V}$. Here, we extend this discussion by including $g_{_N}$ and, for both representations, the associated scalar curvature densities, defined by
\[
\mathcal{R}=\sqrt{|g|}R,
\]
which combines the curvature scalar with the invariant measure of the thermodynamic manifold. To the best of our knowledge, this is the first systematic study of scalar curvature densities in thermodynamic geometry. By incorporating both curvature and local geometric volume, $\mathcal{R}$ provides complementary information on criticality, phase coexistence, and supercritical behavior beyond that encoded in $R$ alone.\\

We then extend the analysis to more realistic systems. First, we consider fluids interacting via a Lennard–Jones potential, described by the equation of state of Johnson \textit{et al.} \cite{Johnson1993}, which provides an accurate parametrization of simulation data over a wide range of thermodynamic conditions, albeit with mean-field critical behavior. We also examine a more recent parametrization due to Thol \textit{et al.} \cite{Thol2015}, but find that higher-order derivatives required for curvature calculations are not sufficiently stable in this representation.\\

To incorporate non-mean-field critical behavior, we additionally analyze a multiparameter equation of state for argon \cite{Tegeler1999}, obtained from fits to experimental thermodynamic data over a wide range of temperatures and
pressures. Unlike the van der Waals and Lennard-Jones descriptions, which are ultimately grounded in mean-field approximations or molecular simulation data, this equation of state is anchored in experimental measurements and reproduces critical exponents consistent with the Ising universality class, while also yielding stable higher-order derivatives suitable for geometric calculations. This makes it particularly valuable as a benchmark, allowing us to directly compare the geometric structures arising from mean-field and non-mean-field descriptions within a unified framework and to assess which features of thermodynamic geometry persist under physically realistic critical behavior.\\

Using these models, we investigate how the choice of metric and geometric invariant affects key features such as the coexistence curve obtained via the $R$-crossing method, the structure of the supercritical region through curvature extrema and associated Widom lines, and the critical scaling of both $R$ and $\mathcal{R}$. Particular emphasis is placed on disentangling the roles of ensemble dependence and geometric invariants, and on identifying which features reflect intrinsic properties of the fluid and which arise from the chosen thermodynamic representation.

\section{Theory}

We consider two thermodynamic metrics derived from the Helmholtz free energy, corresponding to fixed-volume ($V$-metric) and fixed-particle-number ($N$-metric) representations. These metrics provide two distinct geometric descriptions of the thermodynamic state space, depending on the choice of the extensive fluctuating variable. Although both formulations originate from the same thermodynamic potential and therefore contain equivalent thermodynamic information, they generate different geometric structures away from criticality.

\subsection{$V$-metric}
In this subsection, we introduce the thermodynamic metric constructed in the Helmholtz free energy representation by taking volume $(V)$ as a fixed thermodynamic parameter. This metric provides a geometric description of the thermodynamic state space in terms of temperature $(T)$ and particle number $(N)$, and it is given by \cite{Ruppeiner1995, Vega2022,Rodrigo2025}
\begin{equation}
    g_{_{ij}}^{^{_{(V)}}} = \frac{1}{k T}
    \begin{pmatrix}
    -\frac{\partial^2 F}{\partial T^2} & 0 \\
    0 & \frac{\partial^2 F}{\partial N^2}
    \end{pmatrix}
\end{equation}
where $F$ is the Helmholtz free energy and $k$ is the Boltzmann constant. For practical calculations, it is convenient to express the metric in terms of reduced variables, namely the reduced temperature $T_r = \mathcal{A} T$ and the reduced density $\rho_r = \mathcal{B} \rho = \mathcal{B} N/V$ where $\mathcal{A}$ and $\mathcal{B}$ are scaling parameters. Likewise, the Helmholtz free energy can be rewritten as
\begin{equation}
F = \frac{Vk}{\mathcal{A}\mathcal{B}} f_{V}^\prime,
\end{equation}
where $f_{V}^\prime$ is the dimensionless quantity
\begin{equation}
f_{V}^\prime \equiv \rho_r T_r\, a(T_r,\rho_r),
\end{equation}
with $a = F/(NkT)$ being the reduced Helmholtz free energy, as commonly employed in the equation-of-state (EoS) literature.
Thus, the metric elements associated with the second derivative with respect to temperatures take the following form
\begin{equation}
    g_{_{TT}}^{^{(V)}} = \left( \frac{V \mathcal{A}^2}{\mathcal{B}}\right) \, G_{_{TT}}^{^{(V)}} ,
\end{equation}
where we define the reduce metric element as
\begin{equation}
    G_{_{TT}}^{^{(V)}} \equiv -\frac{1}{T_r}\frac{\partial^2 f_{_{V}}^\prime }{\partial T_r^2}.
\end{equation}
Similarly, the second metric element transforms as
\begin{equation}
    g_{_{\rho \rho}}^{^{(V)}} = \left(\frac{\mathcal{B}}{V }\right)\, G_{_{\rho \rho}}^{^{(V)}},
\end{equation}
where the corresponding reduced metric element is defined by
\begin{equation}
     G_{_{\rho \rho}}^{^{(V)}} \equiv \frac{1}{T_r}\frac{\partial^2 f_{_{V}}^\prime }{\partial \rho_r^2}.
\end{equation}
The determinant of the metric becomes
\begin{equation}
    g_{_{V}} = \mathcal{A}^2 \,  G_{_{V}},
\end{equation}
where the reduced determinant is defined as
\begin{equation}
    G_{_{V}} \equiv G_{_{TT}}^{^{(V)}} \, G_{_{\rho \rho}}^{^{(V)}}. 
\end{equation}
Using these reduced quantities, the scalar curvature takes the form
\begin{align}
R^{^{(V)}} &= \left( \frac{\mathcal{B}}{V}\right)\frac{1}{\sqrt{G_{_V}}} 
\frac{\partial}{\partial T_r} \left[
\frac{1}{\sqrt{G_{_V}}} 
\frac{\partial G_{_{\rho \rho}}^{^{(V)}} }{\partial T_r}
\right] \notag \\
       & + \left( \frac{\mathcal{B}}{V}\right) \frac{1}{\sqrt{G_{_V}}} 
\frac{\partial}{\partial \rho_r} \left[
\frac{1}{\sqrt{G_{_V}}} 
\frac{\partial G_{_{TT}}^{^{(V)}} }{\partial \rho_r}
\right]
\end{align}
which can be written in reduced form as
\begin{equation}
    R_{_{V}} \equiv \left( \frac{V}{\mathcal{B}}\right) R^{^{(V)}}.
\end{equation}
\subsection{$N$-metric}
Using the same Helmholtz free energy representation, but now considering the particle number $(N)$ as a fixed thermodynamic parameter, the thermodynamic state space is described by the variables temperature $(T)$ and volume $(V)$, leading to the metric \cite{Ruppeiner1995, Vega2022,Rodrigo2025}
\begin{equation}
    g_{_{ij}}^{^{_{(N)}}} = \frac{1}{k T}
    \begin{pmatrix}
    -\frac{\partial^2 F}{\partial T^2} & 0 \\
    0 & \frac{\partial^2 F}{\partial V^2}
    \end{pmatrix}.
\end{equation}
Using the same rescaling of variables employed in the previous subsection, the Helmholtz free energy can be written as
\begin{equation}
    F = \frac{Nk}{\mathcal{A}} f_{_{N}}^\prime,
\end{equation}
where we have defined the dimensionless quantity,
\begin{equation}
    f_{_{N}}^\prime = T_r a(T_r,\rho_r).
\end{equation}
With this change, the metric elements become
\begin{equation}
    g_{_{TT}}^{^{(N)}} = \left(N \mathcal{A}^2\right) \, G_{_{TT}}^{^{(N)}},
\end{equation}
and
\begin{equation}
    g_{_{\rho \rho}}^{^{(N)}} = \frac{1}{\mathcal{B}^2N}\, G_{_{\rho \rho}}^{^{(N)}},
\end{equation}
where we have defined the reduced metric elements as
\begin{equation}
    G_{_{TT}}^{^{(N)}} \equiv -\frac{1}{T_r}\frac{\partial^2 f_{_{V}}^\prime }{\partial T_r^2},
\end{equation}
and
\begin{equation}
    G_{_{\rho \rho}}^{^{(N)}} \equiv  \frac{1}{T_r} \left( 2\rho_r^3\frac{\partial f_{_{N}}^\prime }{\partial \rho_r}  + \rho_r^4\frac{\partial^2 f_{_{N}}^\prime }{\partial \rho_r^2} \right).
\end{equation}
In the same way, the metric determinant is 
\begin{equation}
    g_{_{N}} =  \left(\frac{\mathcal{A}}{\mathcal{B}} \right)^2 \, G_{_{N}}
\end{equation}
where
\begin{equation}
    G_{_{N}} \equiv  G_{_{TT}}^{^{(N)}} \, G_{_{\rho \rho}}^{^{(N)}},
\end{equation}
with these elements, the scalar curvature in $N$-metric representations is written as
\begin{align}
R^{^{(N)}} &= \left( \frac{1}{N}\right)\frac{1}{\sqrt{G_{_N}}} 
\frac{\partial}{\partial T_r} \left[
\frac{1}{\sqrt{G_{_N}}} 
\frac{\partial G_{_{\rho \rho}}^{^{(N)}} }{\partial T_r}
\right] \notag \\
       & + \left( \frac{1}{N}\right) \frac{\rho_r^2}{\sqrt{G_{_N}}} 
\frac{\partial}{\partial \rho_r} \left[
\frac{\rho_r^2}{\sqrt{G_{_N}}} 
\frac{\partial G_{_{TT}}^{^{(N)}} }{\partial \rho_r}
\right]
\end{align}
that can be reduced as
\begin{equation}
    R_{_{N}} \equiv \left(N \right) R^{^{(N)}}.
\end{equation}
It is important to note that $g_{_{TT}}^{^{(V)}} = g_{_{TT}}^{^{(N)}}$, but $G_{_{TT}}^{^{(V)}} \neq G_{_{TT}}^{^{(N)}}$ since they depend on the functions $f_{_{V}}^\prime$ and $f_{_{N}}^\prime$ respectively.
\subsection{Scalar curvature density}
As mentioned previously, the scalar curvature R is a central quantity in thermodynamic geometry and has been widely used to characterize microscopic interactions, critical behavior, and phase transitions \cite{Ruppeiner2010, Ruppeiner2012, May2013,Branka2018}.

Although $R$ is invariant under coordinate transformations, it is not itself a scalar density and therefore does not incorporate the local volume element of the thermodynamic manifold. 
This naturally motivates the consideration of the geometric density $\sqrt{|g|}R$, which combines the curvature scalar with the invariant measure associated with the metric.\cite{Frankel2011, Wald1984,Nakahara2003} From a geometric perspective, this quantity may provide complementary information by weighting the curvature with the local structure of the thermodynamic state space, potentially revealing aspects of criticality and phase coexistence that are not fully captured by the scalar curvature alone. We define the scalar thermodynamic densities in reduced coordinates as
\begin{equation}
    \mathcal{R}_{_{V}} \equiv \sqrt{|G_{_{V}}|}R_{_{V}},
\end{equation}
\begin{equation}
    \mathcal{R}_{_{N}} \equiv \sqrt{|G_{_{N}}|}R_{_{N}}.
\end{equation}
The scalar curvature $R$ encodes information about the microscopic interaction structure and is known to diverge near the critical point as $R \sim t^{-d\nu}$, where $t = |T-T_c|$ \cite{HU2021}. In this work, we also consider the scalar curvature density $\mathcal{R} \equiv \sqrt{|g|}\,R$, where $g = \det(g_{ij})$. 

To determine its critical behavior, it is sufficient to analyze the scaling of the metric determinant. Although the metrics $g_{_V}$ and $g_{_N}$ correspond to different thermodynamic representations, their leading singular behavior near criticality is governed by the same response functions. In both cases, one diagonal component scales as the constant-volume heat capacity, $g_{TT}\sim C_V$, while the second is controlled by the inverse isothermal compressibility, $g_{XX}\sim \kappa_T^{-1}$, where $X$ denotes the corresponding extensive variable. Since the off-diagonal contributions remain non-singular, the determinant is dominated by the product of the diagonal terms, yielding
\[
\det g \sim C_V \kappa_T^{-1},
\]
up to regular prefactors. Using the standard critical scaling laws $C_V\sim t^{-\alpha}$ and $\kappa_T\sim t^{-\gamma}$, it follows that
\[
\sqrt{|g|}\sim t^{(\gamma-\alpha)/2},
\]
and therefore
\[
\mathcal{R}\sim t^{(\gamma-\alpha)/2-d\nu}.
\]

This exponent can be simplified using standard scaling identities \cite{Stanley1971}. Employing hyperscaling, $d\nu=2-\alpha$, one obtains $(\gamma-\alpha)/2-d\nu=(\gamma+\alpha-4)/2$, while the Rushbrooke relation $\alpha+2\beta+\gamma=2$ further reduces this expression to $-(1+\beta)$. Consequently, both scalar densities, $\mathcal{R}_{_{V}}$ and $\mathcal{R}_{_{N}}$, exhibit the universal critical scaling
\[
\mathcal{R}_{_{V}} \sim \mathcal{R}_{_{N}} \sim t^{-(1+\beta)}.
\]

Remarkably, unlike the scalar curvature itself, whose divergence is governed by the correlation length through $R\sim \xi^d$, the scalar density depends only on the order-parameter exponent $\beta$. This indicates that the factor $\sqrt{|g|}$ effectively reweights the curvature by the local thermodynamic volume element, suppressing the dominant contribution associated with the correlation volume and enhancing the geometric imprint of phase separation. In this sense, $\mathcal{R}$ provides a complementary geometric probe of criticality, more directly sensitive to the emergence of macroscopic order.\\

This result is particularly noteworthy because, although the derivation involves the critical exponents $\alpha$, $\gamma$, and $\nu$, the final scaling depends exclusively on the order parameter exponent $\beta$. This differs from the behavior of the scalar curvature, whose divergence is governed by $d\nu$ and is therefore directly tied to the growth of the correlation length. \\

The emergence of $\beta$ as the sole exponent controlling $\mathcal{R}$ indicates that the geometric density emphasizes a different aspect of criticality, associated with the development of macroscopic order rather than the spatial extent of fluctuations. Since $\beta$ governs the vanishing of the order parameter and determines the shape of the coexistence curve near the critical point, the critical behavior of $\mathcal{R}$ suggests a direct connection with the geometry of phase separation. \\

\subsection{ Extension of $R$-crossing method}
\begin{figure}
\centering
\includegraphics[width=1\columnwidth]{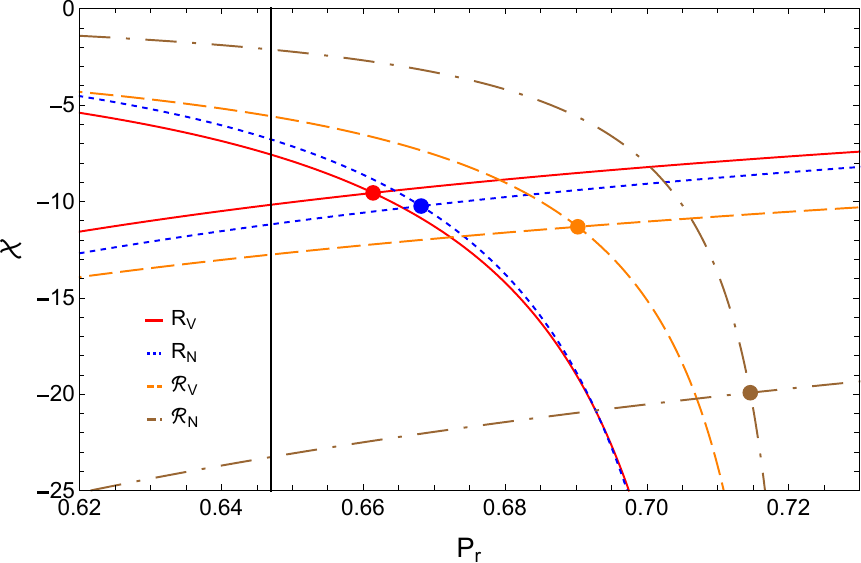}
\caption{\label{Xcrossing}
Crossing of the vapor and liquid branches for the geometrical quantities of the van der Waals fluid at $T_r = 0.90$: $R_{_{V}}$ (red line), $R_{_{N}}$ (dotted blue line), $\mathcal{R}_{_{V}}$ (dashed orange line), and $\mathcal{R}_{_{N}}$ (dash-dotted brown line). The colored points represent the values predicted by the $\mathcal{X}$-crossing method for each quantity, while the vertical black line represents the coexistence vapor-liquid value obtained from the thermodynamic method (black line).
}
\end{figure}

The $R$-crossing method is a procedure that provides a geometric criterion to determine the coexistence of liquid-vapor phases \cite{Ruppeiner1991,Ruppeiner2012}. In this approach, the thermodynamic scalar curvature $R$ is interpreted as a measure of the correlation volume of the system. Near a first-order phase transition, the liquid and vapor phases coexist when the values of the scalar curvature for both phases become equal \cite{Ruppeiner2012}. Therefore, the coexistence condition is obtained from the crossing of the two branches of R, commonly referred to as the R-crossing method. This approach constitutes an alternative to the standard thermodynamic determination of phase coexistence, where the liquid-gas coexistence densities are obtained from the simultaneous solution of the pressure and chemical potential equilibrium conditions. That is, the coexistence densities are determined by solving the following system of equations:
\begin{equation}
    \begin{aligned}
      P(T, \rho_{_{liq}}) -  P(T, \rho_{_{gas}}) &= 0 \\
      \mu(T, \rho_{_{liq}}) -  \mu(T, \rho_{_{gas}}) &= 0 
\end{aligned}
\end{equation}
In the $R$-crossing method, the equilibrium condition associated with the chemical potential is replaced by the requirement that the thermodynamic scalar curvature be equal in the coexisting phases. This procedure has been successfully applied to several fluid systems in two and three dimensions with variable range potentials \cite{JG2019,JG2022,JG2026,Saldana2024,Picon2022}.
In this work, we further propose that an analogous construction can be formulated in terms of the scalar curvature density. Under this assumption, the liquid-gas coexistence is determined by solving
\begin{equation}
    \begin{aligned}
      P(T, \rho_{_{liq}}) -  P(T, \rho_{_{gas}}) &= 0 \\
      \mathcal{X}(T, \rho_{_{liq}}) -  \mathcal{X}(T, \rho_{_{gas}}) &= 0
\end{aligned}
\end{equation}
where $\mathcal{X} = R_{_{V}}, R_{_{N}}, \mathcal{R}_{_{V}}$ or $\mathcal{R}_{_{N}}$. Figure~\ref{Xcrossing} shows the vapor and liquid branches of the four geometric invariants $R_{_V}$, $R_{_N}$, $\mathcal{R}_{_V}$, and $\mathcal{R}_{_N}$ for the van der Waals fluid at $T_r = 0.90$, plotted as a function of reduced pressure. In the $\mathcal{X}$-crossing construction, the coexistence pressure predicted by each invariant is identified as the pressure at which its liquid and gas branches intersect; these predicted values are marked by colored points, while the
vertical black line indicates the coexistence pressure obtained from the standard thermodynamic construction via the simultaneous solution of pressure and chemical potential equilibrium conditions. All four invariants yield crossings in close proximity to the thermodynamic coexistence pressure, confirming that each geometric construction captures the
onset of phase coexistence, while the small but visible quantitative differences among the four predicted values reflect the dependence of the geometric description on both the choice of metric and the specific invariant used. Despite sharing the same underlying thermodynamic potential, the $V$- and $N$-metric representations weight fluctuations differently, and the additional factor $\sqrt{|g|}$ in $\mathcal{R}$ further modifies the location of the branch crossing. That all four constructions remain close to the thermodynamic result demonstrates the robustness of the geometric approach to phase coexistence; the quantitative differences among them, though small at this temperature, grow progressively away from the critical point and provide the primary motivation for the systematic comparison carried out in the subsequent sections.\\

\subsection{Helmholtz free energy for van der Waals fluid}
As a first model, we consider the well-known van der Waals fluid, whose Helmholtz free energy is given by
\begin{equation}\label{EoSvdw}
F = -NkT \left[ \ln\left( \frac{V-Nb}{N\lambda^{3}} \right)+1 \right] - \frac{aN^{2}}{V},
\end{equation}
where $\lambda$ denotes the thermal de Broglie wavelength, $a$ is related to the attractive intermolecular forces, and $b$ to the effective excluded volume of the molecules. For this system, we define the parameters $\mathcal{A}=1/T_c$ and $\mathcal{B}=1/\rho_c$, where the critical temperature and density are
\begin{equation*}
T_c = \frac{8a}{27bk},
\qquad
\rho_c = \frac{1}{3b}.
\end{equation*}
From Eq.~\ref{EoSvdw}, we obtain the functions $f_{V}^{\prime}$ and $f_{N}^{\prime}$, which are required to construct the geometric quantities in both metrics.

\subsection{Helmholtz free energy for Lennard-Jones fluid}\label{sec:models}

The Lennard-Jones potential was employed because it constitutes one of the most representative and widely used models for describing simple fluids with short-range repulsive and long-range attractive interactions. As a first step, we explored the equation of state proposed by Thol \textit{et al.}\
\cite{Thol2015}, which provides an accurate representation of thermodynamic simulation data over a wide range of conditions. However, when computing the thermodynamic curvatures and the associated scalar densities, we found that the higher-order derivatives required for the geometric analysis become
insufficiently stable. In particular, the resulting curvature isotherms exhibited changes in concavity and local structures that were highly sensitive to numerical details and were not reproduced when alternative equations of state were employed. Since the calculation of both $R$ and $\mathcal{R}$ involves derivatives of substantially higher order than those required for standard thermodynamic properties, such behavior suggests that the parametrization, although highly accurate at the level of thermodynamic observables, may not be sufficiently smooth for geometric applications. For this reason, we adopt instead the equation of state of Johnson \textit{et al.}\ \cite{Johnson1993},
which is based on high-precision molecular simulation data, contains $32$ adjustable parameters, and yields stable curvature calculations and a consistent geometric description throughout the thermodynamic region considered. This equation of state accurately reproduces thermodynamic properties over a broad range of temperatures and densities, including the liquid-vapor coexistence region and the critical regime. The Helmholtz free energy is given by

\begin{equation}
    F = F_{id} + F_r
\end{equation}
where $F_{id}$ is the ideal contribution and $F_r$ is the residual part
expressed as
\begin{equation}
   F_r = N \epsilon \left[  \sum_{i=1}^{8}\frac{a_i \rho^{*i}}{i} +
   \sum_{i=1}^{6} b_i G_i \right]
\end{equation}
where $\epsilon$ is the potential well depth, $a_i$ and $b_i$ are temperature dependent coefficients, and $G_i$ are density-dependent coefficients. As for the van der Waals fluid, we define the scaling parameters $\mathcal{A}=1/T_c$
and $\mathcal{B}=1/\rho_c$, where the critical temperature and density are

\begin{equation*}
T_c = 1.313,
\qquad
\rho_c = 0.309,
\end{equation*}
and work throughout with variables reduced by their critical values.
\section{Results}

\subsection{The van der Waals fluid}

\begin{figure}[t]
\centering

\subfloat[Subcritical region.\label{IsoDRsubvdw}]{
    \includegraphics[width=1\columnwidth]{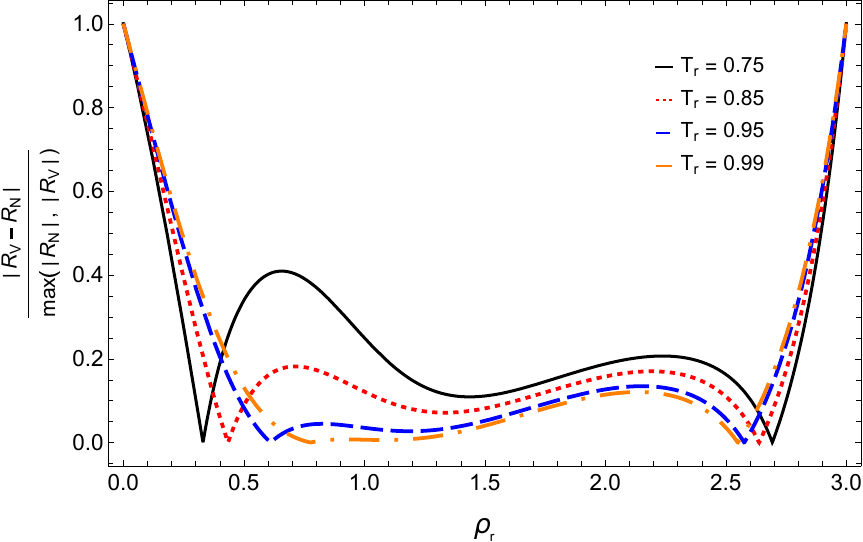}
}

\vspace{0.3cm}

\subfloat[Supercritical region.\label{IsoDRsupvdw}]{
    \includegraphics[width=1\columnwidth]{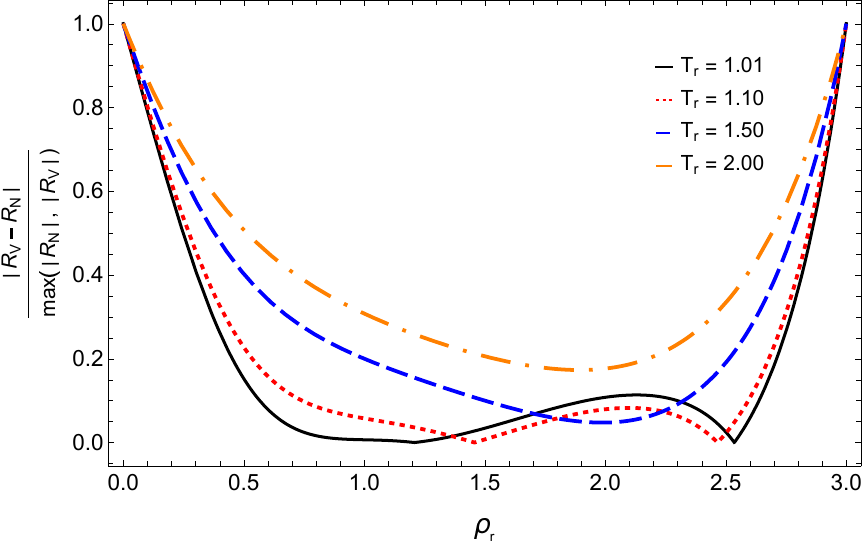}
}

\caption{\label{IsoDRvdw}
Isotherms of the ratio $|R_{_{V}} - R_{_{N}}|/\max ( \vert R_{_{V}}\vert , \vert R_{_{N}} \vert )$ for the van der Waals fluid in the subcritical and supercritical regions.}
\end{figure}

\begin{figure}[t]
\centering

\subfloat[Subcritical region.\label{IsoFDRsubvdw}]{
    \includegraphics[width=1\columnwidth]{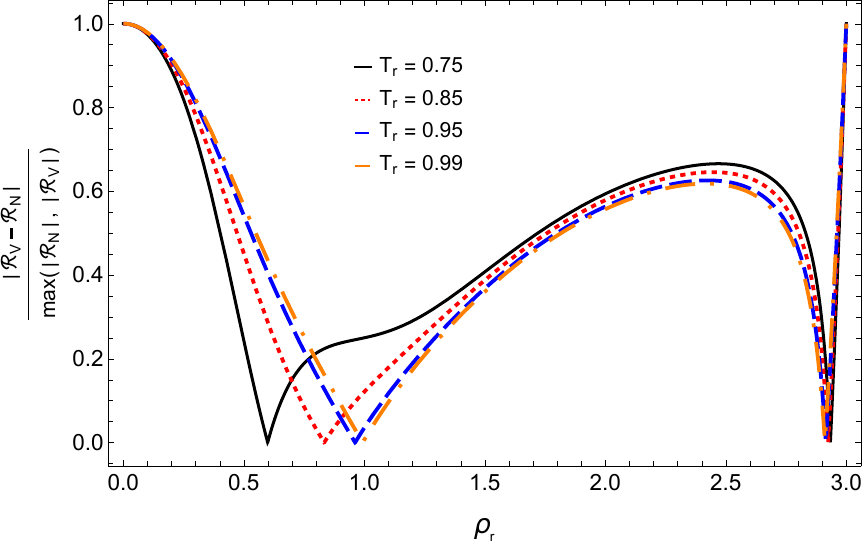}
}

\vspace{0.3cm}

\subfloat[Supercritical region.\label{IsoFDRsupvdw}]{
    \includegraphics[width=1\columnwidth]{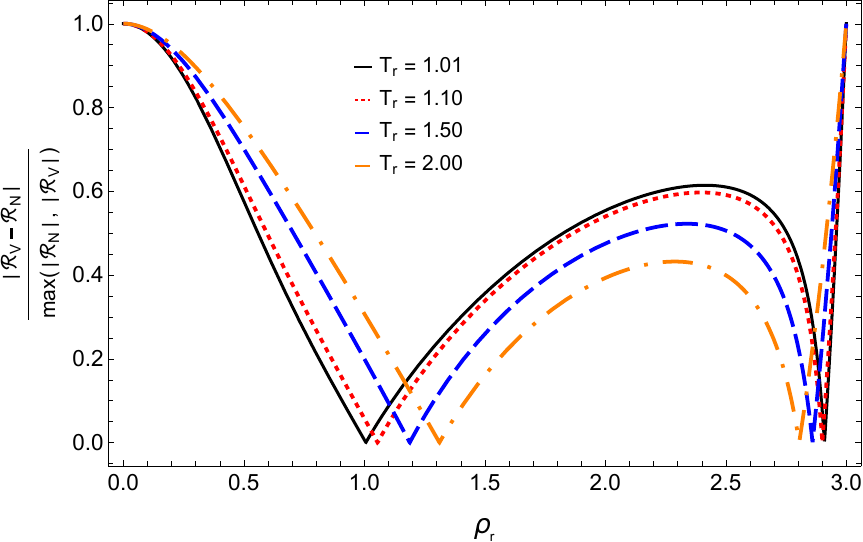}
}

\caption{\label{IsoFDRvdw}
Isotherms of the ratio $|\mathcal{R}_{_{V}} - \mathcal{R}_{_{N}}|/\max(|\mathcal{R}_{_{V}}|,|\mathcal{R}_{_{N}}|)$ for the van der Waals fluid in the subcritical and supercritical regions.}
\end{figure}

\begin{figure}[t]
\centering
\includegraphics[width=1\columnwidth]{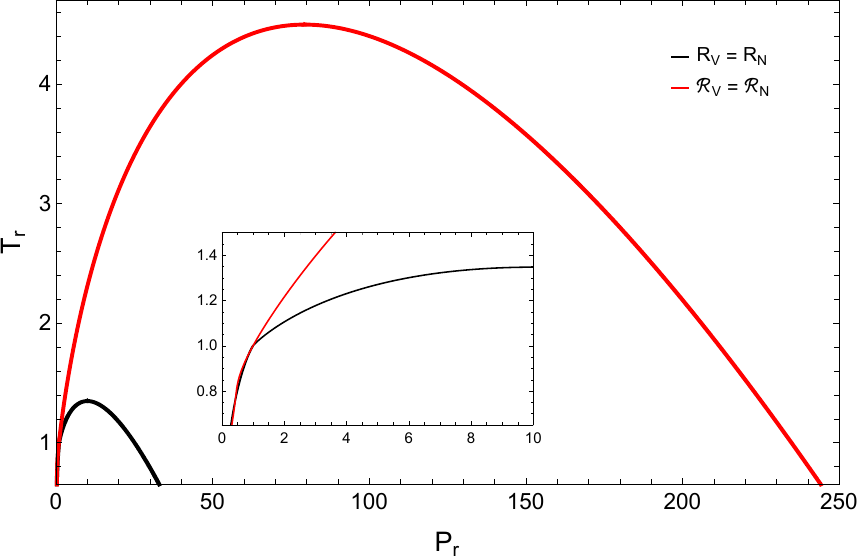}
\caption{\label{IRsvdw}
Thermodynamic states in the $(T,P)$ plane for which the geometric quantities calculated from the $V$-metric and $N$-metric coincide for the van der Waals fluid. The black line corresponds to the condition $R_{_{V}}=R_{_{N}}$, while the red line corresponds to $\mathcal{R}_{_{V}}=\mathcal{R}_{_{N}}$. Along these curves, both thermodynamic representations yield an identical geometric description of the system.
}
\end{figure}

\begin{figure}[t]
\centering

\subfloat[Vapor-liquid coexistence curve in the $(T, P)$ plane.\label{CoexTPvdw}]{
    \includegraphics[width=1\columnwidth]{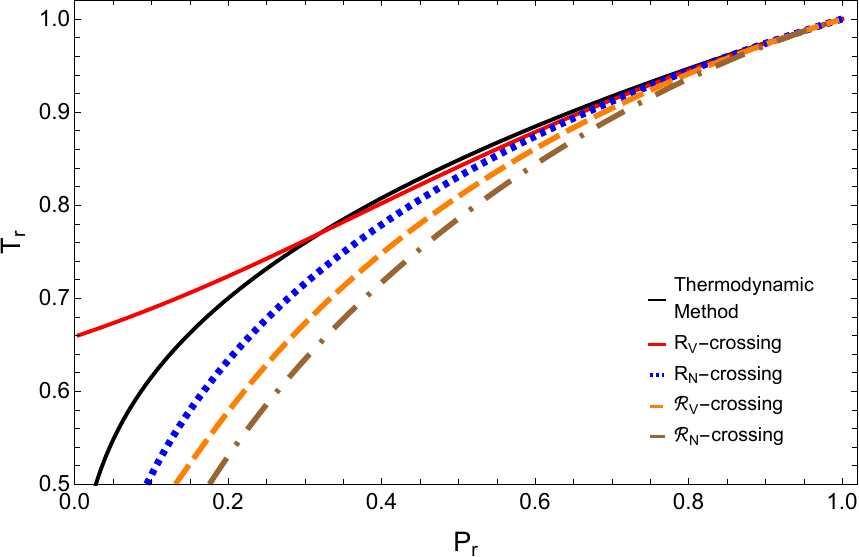}
}

\vspace{0.3cm}

\subfloat[Vapor-liquid coexistence curve in the $(T, \rho)$ plane.\label{CoexTRvdw}]{
    \includegraphics[width=1\columnwidth]{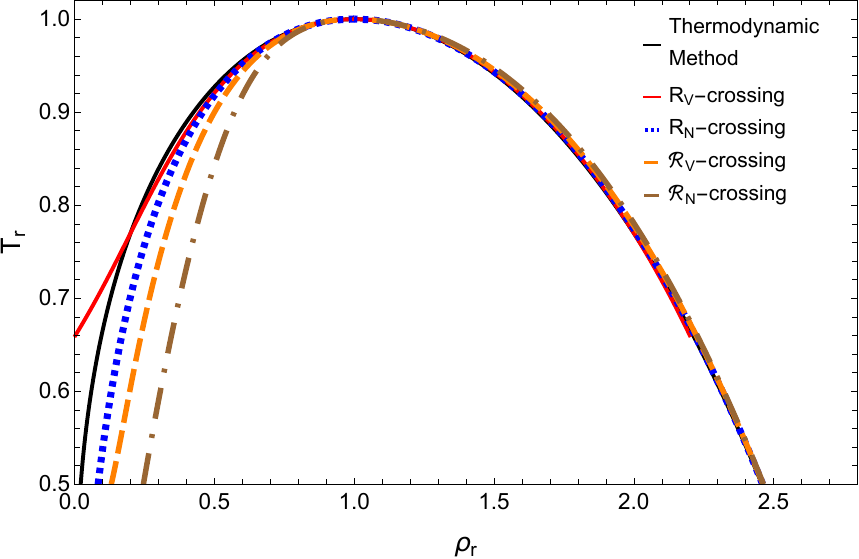}
}

\caption{\label{Coexvdw}
Vapor-liquid coexistence curve for the van der Waals fluid. The black line is calculated using the thermodynamic method, the other ones are obtained using the R-crossing method, with the quantities: $R_{_{V}}$ (red line), $R_{_{N}}$ (dotted blue line), $\mathcal{R}_{_{V}}$ (dashed orange line), and $\mathcal{R}_{_{N}}$ (dash-dotted brown line).}
\end{figure}

\begin{figure}[t]
\centering
\includegraphics[width=1\columnwidth]{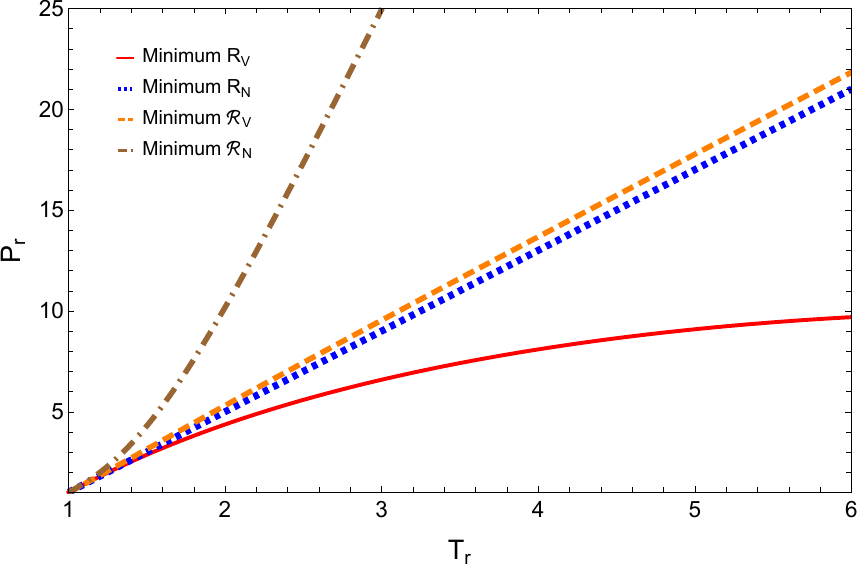}
\caption{\label{Widomvdw}Geometric Widom lines for the van der Waals fluid in the supercritical region,
defined by the loci of minima of the four geometric invariants: $R_{_{V}}$ (red line), $R_{_{N}}$ (dotted blue line), $\mathcal{R}_{_{V}}$ (dashed orange line), and $\mathcal{R}_{_{N}}$ (dash-dotted brown line).}
\end{figure}

We begin our analysis with the van der Waals fluid, which serves as a useful baseline for identifying the general geometric structure of the invariants considered. The standard approach in thermodynamic geometry involves presenting isotherms of the curvature scalar and related quantities in both subcritical and supercritical regimes, where common divergences and characteristic extrema can be observed. However, since the main qualitative features revealed by these isotherms are shared across all models studied here, we focus instead on quantities that allow for a more direct and comparative assessment across models. In particular, we emphasize the behavior of the differences between invariants, the coexistence curves obtained via the generalized $R$-crossing method, the associated Widom lines, and the scaling behavior in the vicinity of the critical point.\\

The relative difference between the scalar curvatures associated with the two metrics is examined,  by plotting isotherms of the quantity $|R_{_{V}} - R_{_{N}}|/max ( \vert R_{_{V}}\vert , \vert R_{_{N}} \vert )$ as a function of the reduced density, see Figures \ref{IsoDRvdw} and \ref{IsoFDRvdw}, where  $\max(f,g)$ denotes the pointwise maximum of the functions $f$ and $g$,  i.e., for each value of the independent variable it returns the larger of $f$ and $g$. At both, low and high-density limits, the ratio approaches unity, indicating maximal deviation between $R_{_V}$ and $R_{_N}$ in these regimes. For a given supercritical temperature, as the density increases from low values, the ratio decreases, reaching one or two well-defined minima depending on the temperature (with two minima appearing at lower temperatures). Beyond these points, the ratio increases again, tending back toward unity at high densities. This behavior reveals that the two curvature scalars become most similar in an intermediate density region, while their differences are amplified in both dilute and dense limits. The emergence of one or two minima suggests a nontrivial restructuring of the geometric response across the supercritical region, likely associated with a crossover behavior.\\

To characterize more precisely the loci where the two geometries become locally indistinguishable, we introduce the \emph{Curvature Equality Curve} (CEC) and the \emph{Curvature-Density Equality Curve} (CDEC). The CEC is defined as the set of thermodynamic states, in the temperature-density plane, at which the scalar curvatures associated with the two metrics coincide,
\begin{equation}
    \mathrm{CEC} \;=\; \Big\{ (T,\rho) \;\Big|\; R_{_{V}}(T,\rho) = R_{_{N}}(T,\rho) \Big\},
\end{equation}
while the CDEC is defined analogously in terms of the scalar curvature densities,
\begin{equation}
    \mathrm{CDEC} \;=\; \Big\{ (T,\rho) \;\Big|\; \mathcal{R}_{_{V}}(T,\rho) = \mathcal{R}_{_{N}}(T,\rho) \Big\},
\end{equation}
with $\mathcal{R}_{_{V}} = \sqrt{|G_{_{V}}|}\,R_{_{V}}$ and $\mathcal{R}_{_{N}} = \sqrt{|G_{_{N}}|}\,R_{_{N}}$. By construction, the CEC corresponds exactly to the zeros of the numerator $|R_{_{V}}-R_{_{N}}|$ appearing in the ratio discussed above, so that the minima identified along each supercritical isotherm mark the points where that isotherm
intersects the CEC. The appearance of one or two such minima therefore reflects how many times a given isotherm crosses the CEC, providing a direct geometric interpretation of the crossover behavior noted previously. We now examine the CEC and CDEC explicitly for the van der Waals fluid, the simplest of the three equations of state considered in this work, before extending the analysis to the Lennard-Jones and argon cases in subsequent sections.\\

Figure~\ref{IRsvdw} shows the CEC and CDEC for the van der Waals fluid in the reduced pressure--temperature plane $(P_r,T_r)$. Both curves originate at low reduced pressure, slightly below the critical temperature, and extend
into the supercritical region, where most of their range lies; only a small portion of each curve corresponds to subcritical states. The CEC ($R_{_{V}}=R_{_{N}}$, black) is confined to a comparatively narrow region at low reduced pressures: after crossing into the supercritical regime, it reaches a shallow maximum and closes back onto the $T_r$ axis. The CDEC ($\mathcal{R}_{_{V}}=\mathcal{R}_{_{N}}$, red), by contrast, spans a much larger region of the phase diagram: it rises to a pronounced maximum before closing onto the $T_r$ axis, nearly an order of magnitude farther than the CEC. As the inset illustrates, the two curves remain close to one another only near their common low-pressure origin, where they appear nearly tangent; beyond $P_r\gtrsim1$ they separate markedly, with the CDEC persisting to states far more removed from the critical region than the CEC. This indicates that the equality of the scalar curvature densities is a substantially less restrictive condition than the equality of the curvature scalars themselves, with the geometric distinction between the two ensembles surviving over a much wider range of thermodynamic states when probed through $R$ alone than when probed through $\mathcal{R}$.\\

We next present the coexistence curves in the $(T,P)$ and $(T,\rho)$ planes, see Figure~\ref{Coexvdw}, obtained from the standard thermodynamic construction and compared with those reconstructed via the $R$-crossing method using $R_{_{V}}$ and $R_{_{N}}$ separately. Although the van der Waals fluid has been extensively studied within the framework of thermodynamic geometry, including analyses based on both the $g_{_{V}}$ and $g_{_{N}}$ metrics, applications of the $R$-crossing method have focused almost exclusively on the curvature associated with $g_{_{V}}$. To the best of our knowledge, a systematic implementation of the $R$-crossing construction using $R_{_{N}}$ has not been reported. In addition, we introduce an extended geometric construction in which phase coexistence is determined by imposing equality of the corresponding scalar densities, $\sqrt{|g|}R$. While such conditions do not follow directly from the original $R$-crossing argument, they provide a natural extension of the geometric approach and allow us to assess whether curvature densities encode coexistence information in a manner analogous to the curvature scalar itself. The van der Waals fluid therefore offers an ideal testing ground for comparing the predictive capabilities of these four geometric invariants under controlled conditions.\\

We find that enforcing equality of any of these four quantities allows, to varying degrees, a reconstruction of the coexistence curve. In the $(T,P)$ plane, the curve obtained from $R_{_{V}}$ remains closest to the thermodynamic coexistence curve up to temperatures nearest the critical point, followed in accuracy by $R_{_{N}}$, then by the density associated with $R_{_{V}}$, and finally by that associated with $R_{_{N}}$. More striking behavior is observed in the $(T,\rho)$ representation: while the same hierarchy is broadly preserved, the curves obtained from $R_{_{N}}$ and from both scalar densities, unlike that from $R_{_{V}}$, maintain the correct qualitative trend over the entire temperature range. Moreover, along the liquid branch, these latter constructions not only capture the correct trend but reproduce the coexistence densities with remarkable accuracy, essentially overlapping with the thermodynamic coexistence curve.
The ability of both scalar curvatures and curvature densities to reconstruct the coexistence curve well beyond the immediate vicinity of the critical point suggests that these geometric quantities retain information relevant to phase coexistence over a much broader region of the thermodynamic manifold than originally anticipated from the $R$-crossing argument. Particularly noteworthy is the behavior along the liquid branch, where all four constructions remain highly accurate even far from criticality. This may indicate that the geometric description becomes less sensitive to the choice of metric representation in this region, or alternatively that the liquid phase is characterized by a more robust geometric structure that is consistently captured by the different invariants considered here.\\

Finally, in Figure~\ref{Widomvdw} we analyze the Widom lines defined from the loci of minima of the four geometric quantities $R_{_{V}}$, $R_{_{N}}$, $\mathcal{R}_{_{V}}$, and $\mathcal{R}_{_{N}}$ in the supercritical region of the van der Waals fluid. Since all four quantities exhibit well-defined minima above the critical temperature, it is possible to construct a corresponding Widom line for each case. As expected, all four lines converge in the vicinity of the critical point, reflecting the common critical behavior shared by the underlying geometric invariants. Away from criticality, however, clear differences emerge. The Widom lines associated with $R_{_{N}}$ and $\mathcal{R}_{_{V}}$ are remarkably similar, displaying an almost linear behavior that extends far into the supercritical region. In contrast, the line defined by $R_{_{V}}$ lies below these two and exhibits a concave-downward shape, while the line associated with $\mathcal{R}_{_{N}}$ has the largest slope, rapidly separating from the others and showing an opposite (concave-upward) curvature. Interestingly, the two intermediate lines appear nearly straight over a wide temperature range. The existence of four distinct Widom lines demonstrates that the geometric characterization of the supercritical region depends not only on the choice of metric but also on the choice of geometric invariant. While all constructions recover the same critical limit, their progressive separation away from the critical point indicates that different geometric quantities emphasize different aspects of the thermodynamic manifold once critical scaling ceases to dominate the behavior of the system.\\

\begin{figure}[t]
\centering

\subfloat[Subcritical region.\label{IsoDRsubLJ}]{
    \includegraphics[width=1\columnwidth]{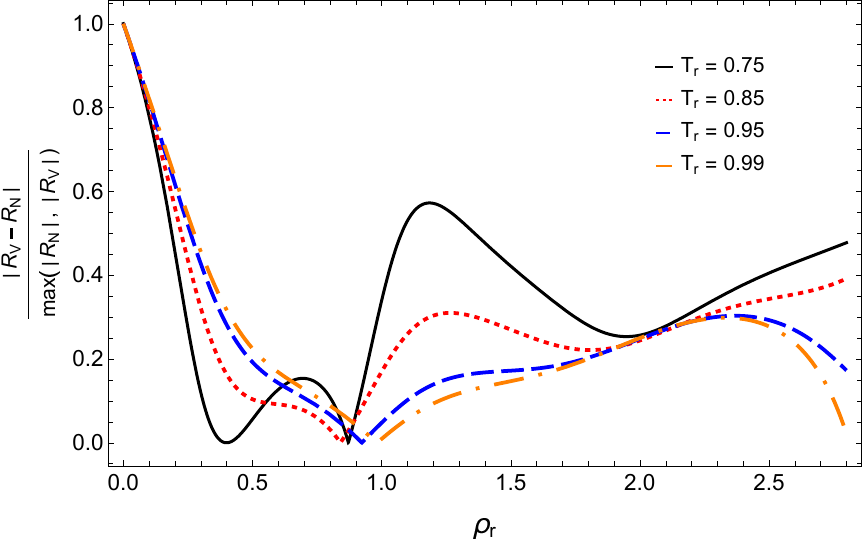}
}

\vspace{0.3cm}

\subfloat[Supercritical region.\label{IsoDRsupLJ}]{
    \includegraphics[width=1\columnwidth]{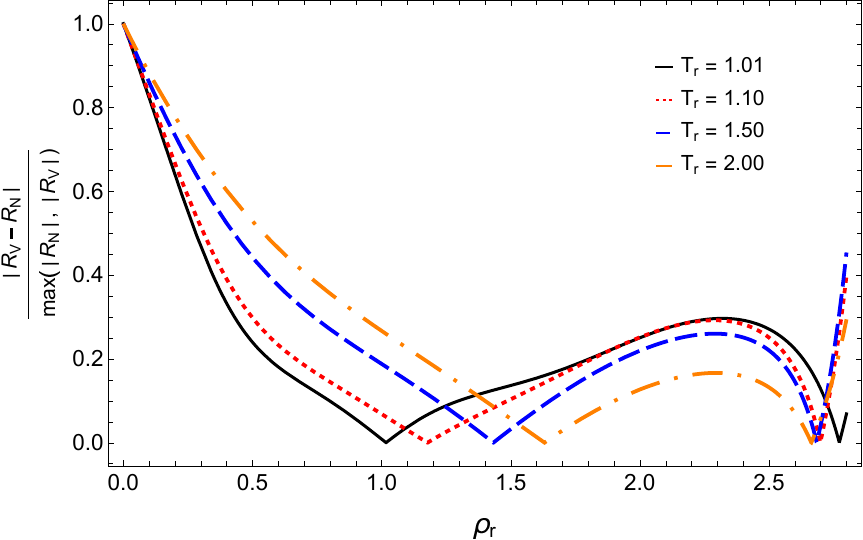}
}

\caption{\label{IsoDRLJ}
Isotherms of the ratio $|R_{_{V}} - R_{_{N}}|/max ( \vert R_{_{V}}\vert , \vert R_{_{N}} \vert )$ for the Lennard-Jones fluid in the subcritical and supercritical regions.}
\end{figure}

\begin{figure}[t]
\centering

\subfloat[Subcritical region.\label{IsoFDRsubLJ}]{
    \includegraphics[width=1\columnwidth]{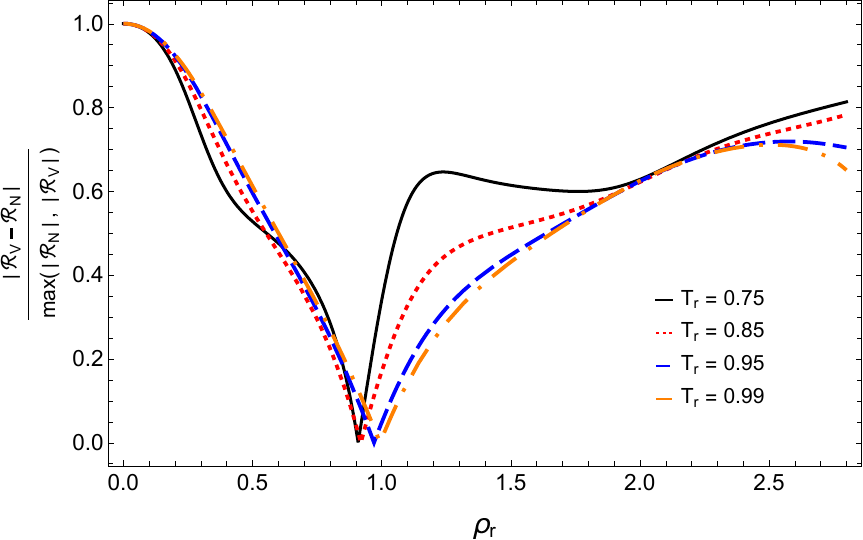}
}

\vspace{0.3cm}

\subfloat[Supercritical region.\label{IsoFDRsupLJ}]{
    \includegraphics[width=1\columnwidth]{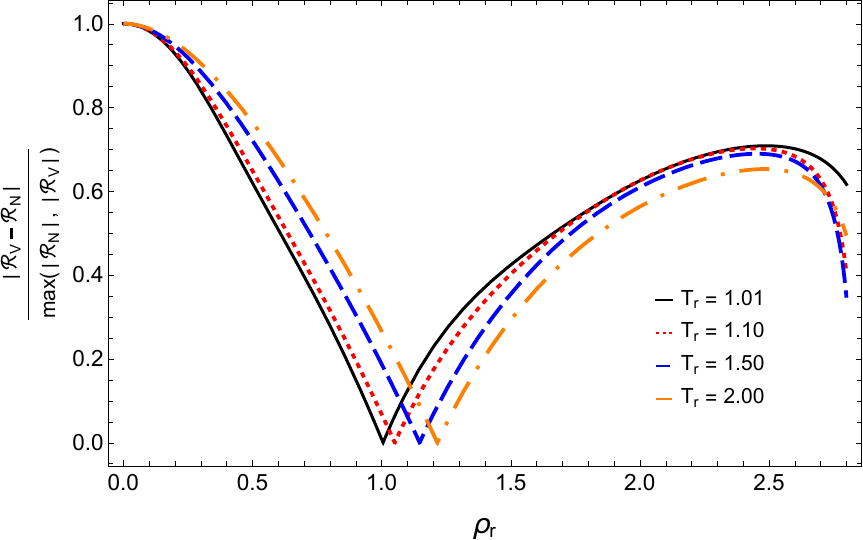}
}

\caption{\label{IsoFDRLJ}
Isotherms of the ratio $|\mathcal{R}_{_{V}} - \mathcal{R}_{_{N}}|/\max(|\mathcal{R}_{_{V}}|,|\mathcal{R}_{_{N}}|)$ for the Lennard-Jones fluid in the subcritical and supercritical regions.}
\end{figure}

\begin{figure}[t]
\centering
\includegraphics[width=1\columnwidth]{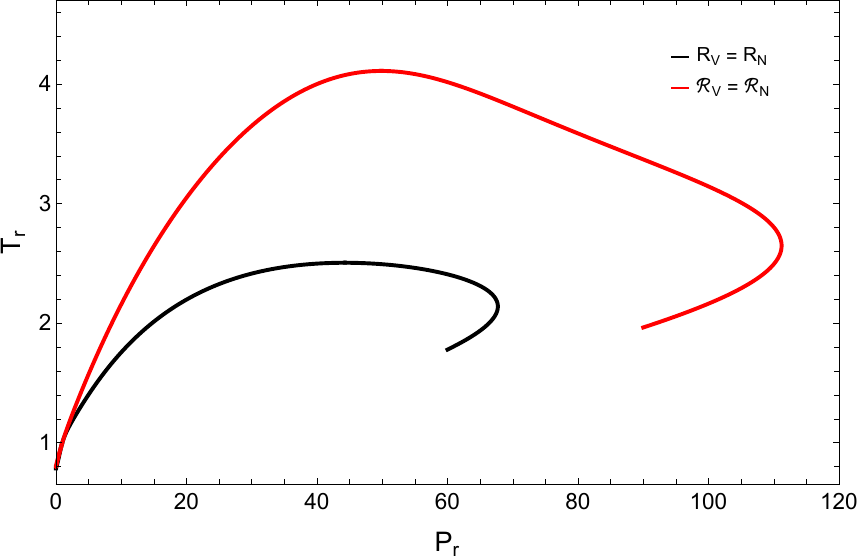}
\caption{\label{IRsLJ}
Thermodynamic states in the $(T,P)$ plane for which the geometric quantities calculated from the $V$-metric and $N$-metric coincide for the Lennard-Jones fluid. The black line corresponds to the condition $R_{_{V}}=R_{_{N}}$, while the red line corresponds to $\mathcal{R}_{_{V}}=\mathcal{R}_{_{N}}$. Along these curves, both thermodynamic representations yield an identical geometric description of the system.
}
\end{figure}

\begin{figure}[t]
\centering

\subfloat[Vapor-liquid coexistence curve in the $(T, P)$ plane.\label{CoexTPLJ}]{
    \includegraphics[width=1\columnwidth]{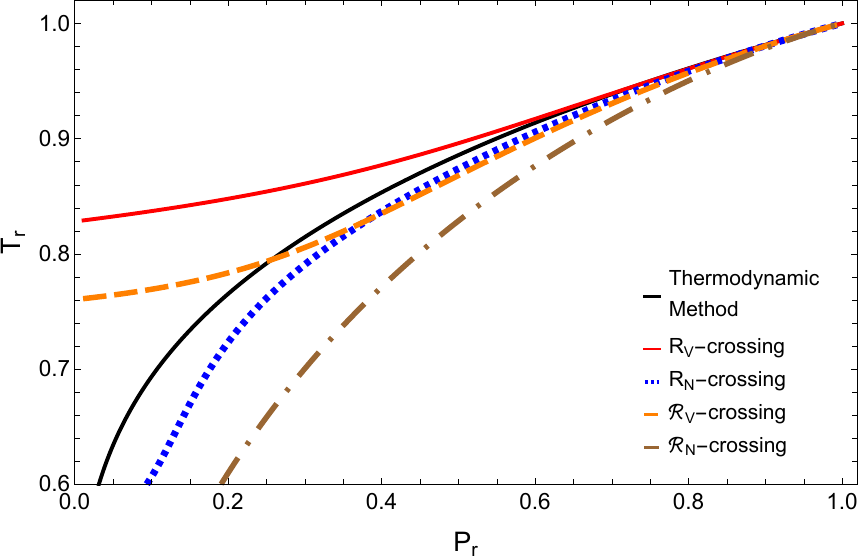}
}

\vspace{0.3cm}

\subfloat[Vapor-liquid coexistence curve in the $(T, \rho)$ plane.\label{CoexTRLJ}]{
    \includegraphics[width=1\columnwidth]{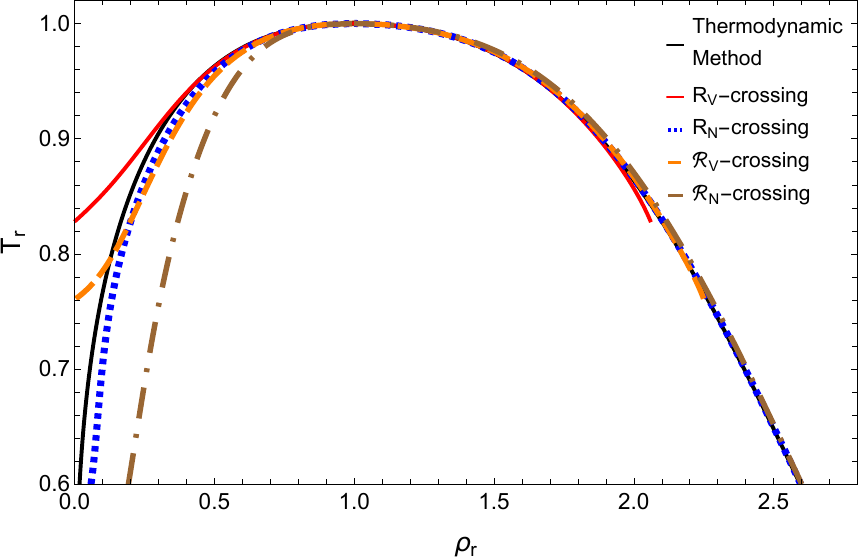}
}

\caption{\label{CoexLJ}
Vapor-liquid coexistence curve for the Lennard-Jones fluid. The black line is calculated using the thermodynamic method, the other ones are obtained using the R-crossing method, with the quantities: $R_{_{V}}$ (red line), $R_{_{N}}$ (dotted blue line), $\mathcal{R}_{_{V}}$ (dashed orange line), and $\mathcal{R}_{_{N}}$ (dash-dotted brown line).}
\end{figure}

\begin{figure}[t]
\centering
\includegraphics[width=1\columnwidth]{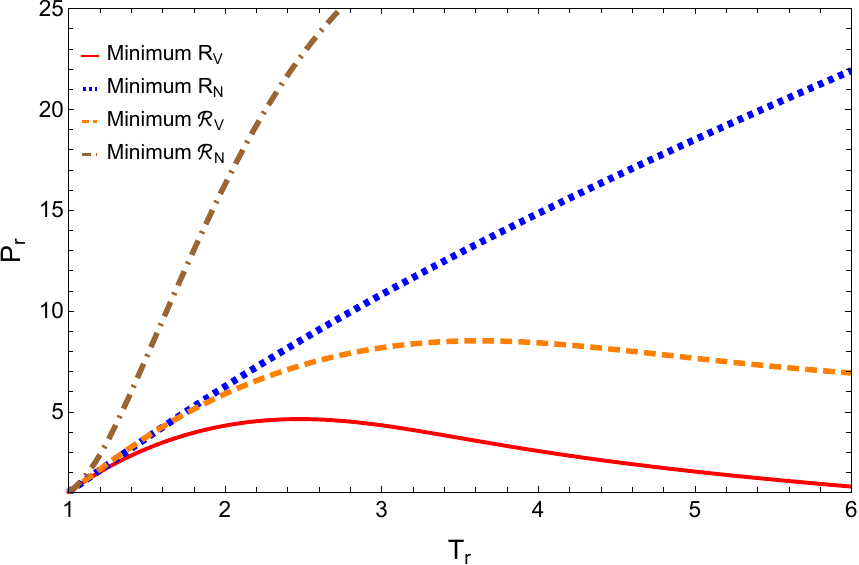}
\caption{Geometric Widom lines for the Lennard-Jones fluid in the supercritical region, defined by the loci of minima of the four geometric invariants: $R_{_{V}}$ (red line), $R_{_{N}}$ (dotted blue line), $\mathcal{R}_{_{V}}$ (dashed orange line), and $\mathcal{R}_{_{N}}$ (dash-dotted brown line).} \label{WidomLJ}
\end{figure}

\subsection{The Lennard-Jones fluid}

We now extend the analysis performed for the van der Waals fluid to a more realistic Lennard-Jones (LJ) fluid, described by the equation of state of Johnson \textit{et al.}\ \cite{Johnson1993}, which yields stable higher-order derivatives and a consistent geometric description throughout the thermodynamic region considered. As discussed in Section~\ref{sec:models}, this choice is motivated by the numerical requirements of geometric calculations, where derivatives of substantially higher order than those needed for standard thermodynamic properties must be evaluated reliably.\\

The relative difference between the scalar curvatures associated with the two metrics is examined by plotting isotherms of the quantity $|R_{_{V}} - R_{_{N}}|/\max(|R_{_{V}}|,|R_{_{N}}|)$ as a function of the reduced density, see Figures~\ref{IsoDRLJ} and \ref{IsoFDRLJ}. The two curvatures become most similar within an intermediate density range, while their differences increase significantly at both low and high densities. This indicates that the effective equivalence between the metrics is restricted to moderately dense states, where the thermodynamic response is smoother and less sensitive to the choice of representation. In contrast, in the dilute gas and in the high-density liquid regimes, the geometric structures induced by $g_{_{V}}$ and $g_{_{N}}$ diverge more strongly, leading to larger discrepancies in the curvature scalars. The approach of the ratio to unity in these limits reflects this growing disparity.\\

This behavior highlights that different metric constructions probe distinct aspects of the thermodynamic response once the system moves away from the intermediate regime. In particular, the region where both metrics agree most
closely coincides with the domain where the fluid exhibits the most balanced interplay between density and energetic fluctuations, whereas deviations appear as either density fluctuations (in the dilute limit) or local structural rigidity (in the dense limit) become dominant. A similar structure was already observed in the van der Waals fluid, where the relative difference between $R_{_{V}}$ and $R_{_{N}}$ also exhibited a minimum at intermediate densities and increased towards both low- and high-density limits. The persistence of this pattern in the Lennard-Jones fluid suggests that this behavior is not specific to the mean-field character of vdW, but may instead reflect a more general feature of thermodynamic geometry: namely, that the equivalence between ensemble-dependent metrics emerges only in a restricted region of the thermodynamic manifold where no single fluctuation channel dominates the response.\\

Figure~\ref{IRsLJ} shows the CEC and CDEC for the Lennard-Jones fluid in the reduced pressure--temperature plane. The overall topology is qualitatively similar to that observed for the van der Waals fluid, though quantitative differences are apparent, reflecting the more realistic molecular interaction framework underlying the Lennard-Jones equation of state. As in the van der Waals case, both curves originate slightly below the critical temperature at low reduced pressures and extend predominantly into the supercritical region, with the CDEC spanning a substantially larger region of the phase diagram than the CEC. This confirms that the equality of scalar curvature densities is a less restrictive condition than the equality of curvature scalars, a pattern that persists beyond the mean-field limit.\\

We begin with the coexistence curves constructed via the $R$-crossing method using $R_{_{V}}$, $R_{_{N}}$, and their associated scalar densities. In the $(T,P)$ plane, Figure~\ref{CoexTPLJ}, a behavior similar to that of the van der Waals fluid is observed, with the important difference that not only the curve obtained from $R_{_{V}}$, but also that constructed from $\mathcal{R}_{_{V}}$, deviates significantly from the thermodynamic coexistence curve as one moves away from the critical point. In contrast, the curves obtained from $R_{_{N}}$ and $\mathcal{R}_{_{N}}$ preserve the correct qualitative trend, although they also exhibit a gradual deviation at lower temperatures, which remains comparatively moderate. As in the van der Waals case, the behavior is more clearly resolved in the $(T,\rho)$ plane, Figure~\ref{CoexTRLJ}. There, both $R_{_{V}}$ and its associated density show a pronounced departure from the coexistence curve as the system moves away from criticality, but this discrepancy is essentially confined to the gas branch, while the liquid branch remains well described. More notably, the coexistence
curve reconstructed from $R_{_{N}}$ closely follows the thermodynamic result in the gas phase and becomes nearly exact along the liquid branch. Indeed, as in the van der Waals case, all four constructions yield highly accurate results in the liquid phase, even far from the critical point. The improved performance of $R_{_{N}}$ and its associated density in reconstructing the coexistence curve is consistent with the behavior of the ratio discussed above, where the $N$-metric and $V$-metric agree most closely precisely in the intermediate density regime that is most relevant for phase coexistence.\\

Figure~\ref{WidomLJ} shows the geometric Widom lines for the Lennard-Jones fluid, defined by the loci of minima of the four invariants $R_{_{V}}$, $R_{_{N}}$, $\mathcal{R}_{_{V}}$, and $\mathcal{R}_{_{N}}$ in the supercritical region. All four lines converge at the critical point, reflecting the common critical behavior shared by the underlying geometric invariants. Away from criticality, however, they separate progressively and display qualitatively distinct behaviors, reproducing the same fourfold structure identified for the van der Waals fluid. The Widom lines associated with $R_{_{N}}$ and $\mathcal{R}_{_{V}}$ are remarkably similar, displaying an almost linear extension into the supercritical region. In contrast, the line defined by $R_{_{V}}$ exhibits a concave-downward shape, while that associated with $\mathcal{R}_{_{N}}$ has the largest slope, rapidly separating from the others with a concave-upward curvature. The persistence of this fourfold structure in the Lennard-Jones fluid confirms that the metric and invariant dependence of the supercritical geometric characterization is not specific to
the van der Waals model but survives in a more realistic fluid description.\\

Finally, the critical behavior of the four geometric quantities was analyzed by fitting the critical scaling form
\begin{equation}
\ln |\mathcal{X}| = -a^\prime \ln|1-T_r| + b^\prime,
\end{equation}
near the critical point. The density was fixed at its critical value, while the reduced temperature was approached from the supercritical region, $T_r>1$. Table~\ref{Critval} summarizes the corresponding critical exponents. The exponents obtained for the Lennard-Jones fluid, described by the Johnson equation of state, are consistent with the expected mean-field values rather than those of the three-dimensional Ising universality class. This behavior should not be interpreted as a property of the Lennard-Jones fluid itself, whose asymptotic critical behavior is known to belong to the Ising universality class, but rather as a limitation of the underlying equation of state. Although the Johnson equation provides an accurate representation of molecular simulation data over a broad range of thermodynamic conditions, the simulations on which it is based do not incorporate the finite-size scaling analysis required to recover the true asymptotic critical behavior. Consequently, the resulting parametrization reproduces an effective mean-field critical regime in the immediate vicinity of the critical point. In contrast, the multiparameter equation of state for argon, constructed from high-accuracy experimental data and designed to reproduce the correct asymptotic critical behavior, yields exponents in close agreement with those of the three-dimensional Ising universality class. These results provide an additional validation of the geometric framework and confirm that the critical behavior of both the scalar curvature and the scalar density faithfully reflects the universality class encoded in the underlying thermodynamic description.

\begin{figure}[t]
\centering
\includegraphics[width=1\columnwidth]{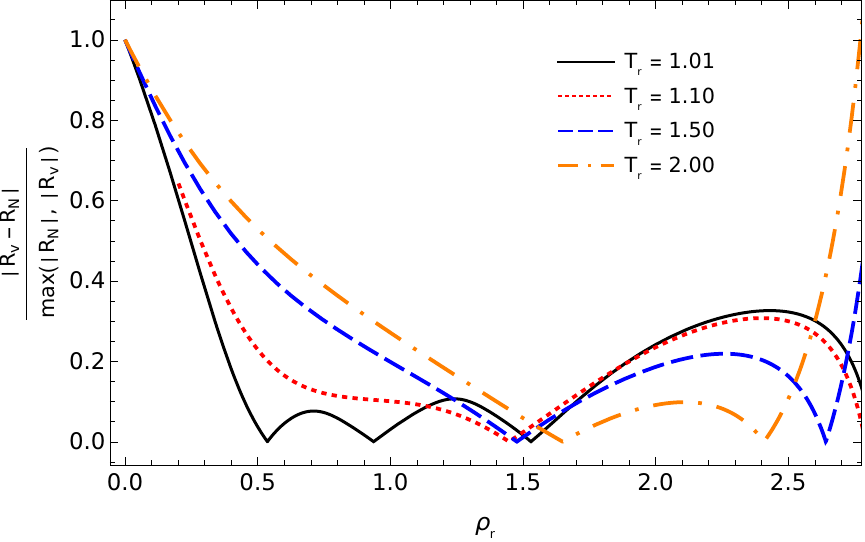}
\caption{\label{IsoDifAr}Isotherms of the ratio $|R_{_{V}} - R_{_{N}}|/max ( \vert R_{_{V}}\vert , \vert R_{_{N}} \vert )$ for Argon in the supercritical region.}
\end{figure}

\begin{figure}[t]
\centering
\includegraphics[width=1\columnwidth]{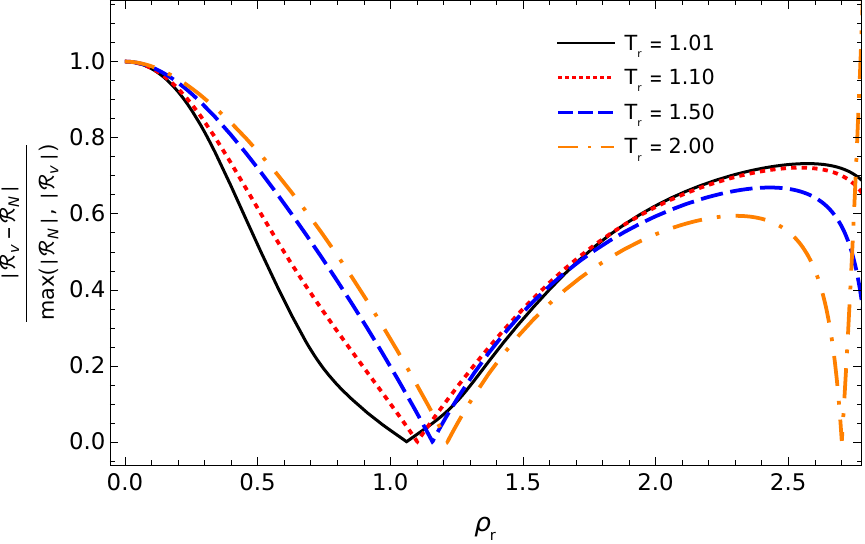}
\caption{\label{IsoFDifAr}Isotherms of the ratio $|\mathcal{R}_{_{V}} - \mathcal{R}_{_{N}}|/\max(|\mathcal{R}_{_{V}}|,|\mathcal{R}_{_{N}}|)$ for Argon fluid in supercritical region.}
\end{figure}

\begin{figure}[t]
\centering
\includegraphics[width=1\columnwidth]{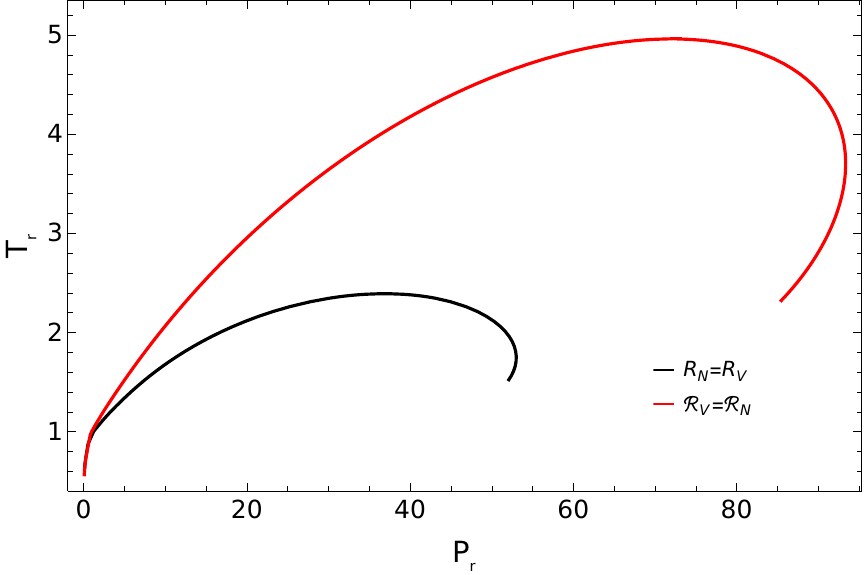}
\caption{Thermodynamic states in the $(T,P)$ plane for which the geometric quantities calculated from the $V$-metric and $N$-metric coincide for Argon. The black line corresponds to the condition $R_{_{V}}=R_{_{N}}$, while the red line corresponds to $\mathcal{R}_{_{V}}=\mathcal{R}_{_{N}}$. Along these curves, both thermodynamic representations yield an identical geometric description of the system.
\label{IRsAr}
}
\end{figure}

\begin{figure}[t]
\centering

\subfloat[Vapor-liquid coexistence curve in the $(T, P)$ plane.\label{CoexTPAr}]{
    \includegraphics[width=1\columnwidth]{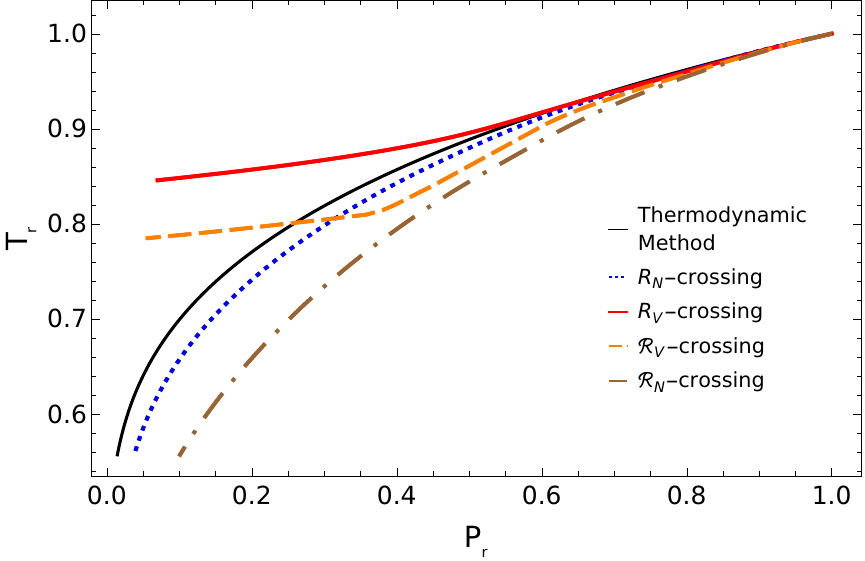}
}

\vspace{0.3cm}

\subfloat[Vapor-liquid coexistence curve in the $(T, \rho)$ plane.\label{CoexTRAr}]{
    \includegraphics[width=1\columnwidth]{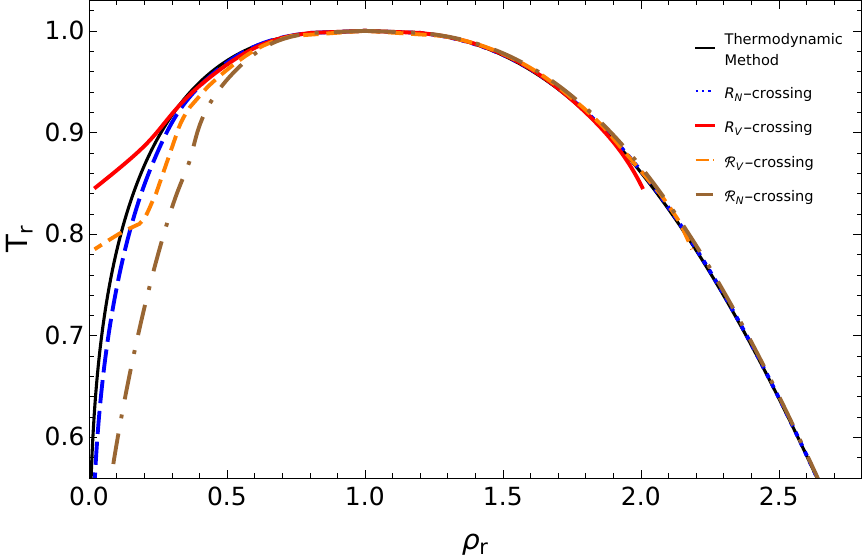}
}

\caption{\label{CoexAr}
Vapor-liquid coexistence curve for Argon. The black line is calculated using the thermodynamic method, the other ones are obtained using the R-crossing method, with the quantities: $R_{_{V}}$ (red line), $R_{_{N}}$ (dotted blue line), $\mathcal{R}_{_{V}}$ (dashed orange line), and $\mathcal{R}_{_{N}}$ (dash-dotted brown line).}
\end{figure}

\begin{figure}[t]
\centering
\includegraphics[width=1\columnwidth]{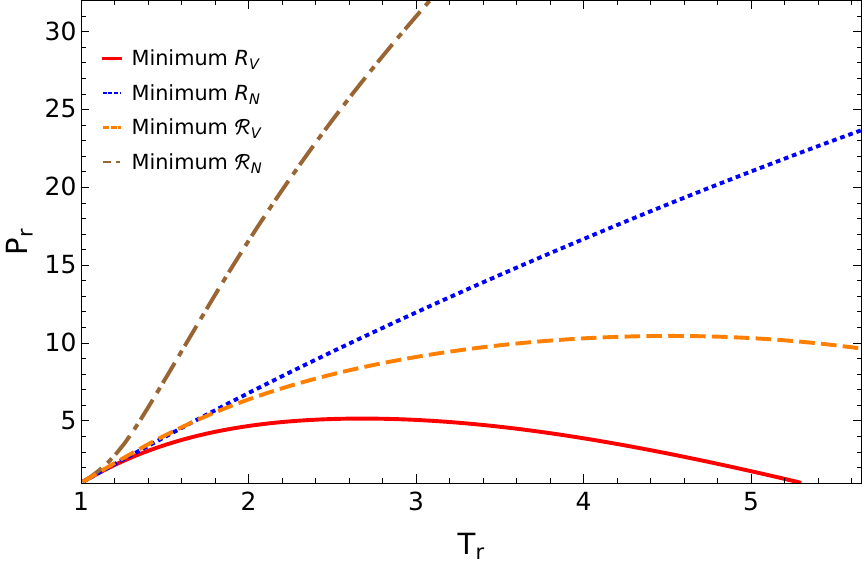}
\caption{\label{WidomAr}
Geometric Widom lines for argon in the supercritical region, defined by the loci of minima of the four geometric invariants: $R_{_{V}}$ (red line), $R_{_{N}}$ (dotted blue line), $\mathcal{R}_{_{V}}$ (dashed orange line), and $\mathcal{R}_{_{N}}$ (dash-dotted brown line).}
\end{figure}

\subsection{The Argon case}

\begin{table}
\caption{\label{Critval}
Parameters obtained from fitting the critical scaling relation $\ln |\mathcal{X}| = -a^\prime \ln|1-T_r| + b^\prime$ for the geometrical quantities near the critical point.}
\begin{ruledtabular}
\begin{tabular}{lcccc}
 & \multicolumn{2}{c}{Lennard-Jones} & \multicolumn{2}{c}{Argon} \\
 & $a^\prime$ & $b^\prime$ & $a^\prime$ & $b^\prime$\\
\hline
$R_{_{V}}$         & $1.998622$ & $-1.454719$ & 1.88428 & -1.42336 \\
$R_{_{N}}$         & $1.998962$ & $-1.457816$ & 1.89266 & -1.48786 \\
$\mathcal{R}_{_{V}}$ & $1.499331$ & $-0.619466$ & 1.32536 & -0.193931 \\
$\mathcal{R}_{_{N}}$ & $1.499671$ & $-0.622564$ & 1.33976 & -0.294097 \\
\end{tabular}
\end{ruledtabular}
\end{table}
To further strengthen the analysis and assess the generality of the observed geometric features beyond both mean-field models and equations of state derived from molecular simulations, we next consider a multiparameter equation of state for argon fitted directly to experimental data. While the van der Waals fluid provides a classical mean-field reference system, and the Lennard--Jones description based on simulation-informed equations of state introduces a more realistic molecular interaction framework, neither is ultimately grounded in experimental  behavior. In contrast, the argon equation of state incorporates multiparameter fits to experimental thermodynamic data and is known to correctly capture non-mean-field critical scaling consistent with the Ising universality class. This makes it particularly suitable as a benchmark to test the extent to which the geometric structures identified in the previous sections persist under physically realistic critical behavior.\\

In Figure~\ref{IsoDifAr} isotherms of the normalized difference $|R_{_{V}} - R_{_{N}}|/\max(|R_{_{V}}|,|R_{_{N}}|)$ as a function of the reduced density for argon at four supercritical temperatures are shown. The overall behavior closely mirrors that observed for the van der Waals and Lennard-Jones fluids: the ratio approaches unity in both the low- and high-density limits, reflecting maximal geometric disagreement between the two metrics in these regimes, while one or two well-defined minima, where the ratio vanishes exactly, appear at intermediate densities. The isotherm nearest the critical point exhibits two such zeros, at low and high intermediate densities, corresponding to two distinct intersections with the CEC. At higher temperatures, only a single zero is observed, indicating that these isotherms cross the CEC only once. This transition from two zeros to one as the temperature increases above the
critical point is consistent with the structure of the CEC discussed previously, and confirms that the crossover behavior identified in the simpler fluid models persists in argon, whose equation of state correctly captures
non-mean-field critical behavior consistent with the Ising universality class.\\

Analogously, Figure~\ref{IsoFDifAr} shows the corresponding isotherms of the normalized difference $|\mathcal{R}_{_{V}} - \mathcal{R}_{_{N}}|/ \max(|\mathcal{R}_{_{V}}|,|\mathcal{R}_{_{N}}|)$. The overall behavior is
similar to that previously observed for the Lennard-Jones fluid. In this case, the normalized difference exhibits a zero located at nearly the same reduced density for all the supercritical temperatures considered. For both invariants, only the supercritical region is shown. In the subcritical region, the evaluation of the geometric quantities becomes numerically unstable, manifesting as loss of smoothness in the curvature isotherms. This instability is absent in the supercritical region, where the calculations remain well-behaved throughout. The precise origin of this behavior is not fully understood; however, we attribute it tentatively to the same class of difficulty encountered with the Thol equation of state for the Lennard-Jones fluid, namely that the multiparameter fitting procedure, while highly accurate at the level of standard thermodynamic observables, may not guarantee sufficient smoothness in higher-order derivatives across the entire fluid region. The subcritical domain, where the thermodynamic response varies more sharply in the vicinity of the coexistence curve, may be particularly susceptible to such limitations.\\

CEC and CDEC for argon in the reduced pressure-temperature plane are shown in Figure~\ref{IRsAr}. The overall topology is qualitatively similar to that observed for the Lennard-Jones fluid, and differs markedly from the van
der Waals case, reinforcing the conclusion that the geometric structure of these curves is more sensitive to the realism of the equation of state than to the universality class at the critical point. Both curves originate slightly
below the critical temperature at low reduced pressures and extend predominantly into the supercritical region. The CEC ($R_{_{V}} = R_{_{N}}$, black) rises to a moderate maximum before closing back on itself, remaining confined
to a comparatively limited region of the phase diagram. The CDEC ($\mathcal{R}_{_{V}} = \mathcal{R}_{_{N}}$, red) spans a substantially larger region, reaching a pronounced maximum at considerably higher temperatures and pressures
before descending steeply. Both curves terminate abruptly rather than closing smoothly onto the $T_r$ axis, as was the case for the van der Waals fluid; this termination reflects the boundary of the thermodynamic domain covered by
the argon equation of state. Within the accessible domain, the CDEC encloses an area roughly an order of magnitude larger than the CEC, consistent with the pattern established for the other two fluid models and confirming that the
equality of scalar curvature densities is a substantially less restrictive condition than the equality of curvature scalars across the thermodynamic manifold.\\

Figure~\ref{CoexAr} shows the vapor--liquid coexistence curve for argon reconstructed via the generalized $R$-crossing method using all four geometric invariants, compared against the thermodynamic result in both the $(T_r, P_r)$  and $(T_r, \rho_r)$ planes. The overall pattern is qualitatively consistent with that observed for the van der Waals and Lennard-Jones fluids, though the reconstruction is numerically more demanding due to the complex parametrization of the argon equation of state. In the $(T_r, P_r)$ plane, all four constructions converge to the thermodynamic coexistence curve in the vicinity of the critical point, but diverge progressively as the temperature decreases. Among them, $R_{_{N}}$ provides the closest agreement with the thermodynamic result over the widest temperature range, while $R_{_{V}}$ departs most strongly, lying systematically above the coexistence curve throughout the subcritical region. The constructions based on the scalar curvature densities $\mathcal{R}_{_{V}}$ and $\mathcal{R}_{_{N}}$ fall between these extremes but show considerable deviation at low pressures. In the $(T_r, \rho_r)$ plane, the picture is more revealing: along the liquid branch, all four invariants yield results in close agreement with the thermodynamic coexistence curve even far from the critical point, reproducing the same robustness of the liquid-branch description already noted for the simpler fluid models. Along the vapor branch, however, clear differences emerge: $R_{_{N}}$ again performs best, maintaining good agreement over a wide temperature
range, while $R_{_{V}}$ overestimates the coexistence temperatures and both scalar curvature densities deviate substantially at low densities.\\ 

Finally, the four Widom lines defined by the minima of $R_{_{V}}$, $R_{_{N}}$, $\mathcal{R}_{_{V}}$, and $\mathcal{R}_{_{N}}$ in the supercritical region of argon are presented in Figure~\ref{WidomAr}. As expected, all four lines converge at the critical point, reflecting the common critical behavior shared by the underlying geometric
invariants. Away from criticality, however, they separate markedly and display qualitatively distinct behaviors, reproducing the same fourfold structure identified for the van der Waals and Lennard-Jones fluids. The Widom line
defined by $R_{_{N}}$ displays a concave-downward shape, rising to a moderate maximum before turning back toward lower pressures as the temperature increases. That of $\mathcal{R}_{_{V}}$ behaves similarly but lies at somewhat
higher pressures, also exhibiting a concave-downward profile that eventually flattens and turns over. In contrast, the Widom line associated with $R_{_{V}}$ grows in an approximately linear fashion, extending to substantially higher
pressures without any sign of saturation within the accessible domain. The most striking behavior is that of $\mathcal{R}_{_{N}}$, whose Widom line rises with a markedly superlinear slope, separating rapidly from the other three and reaching the highest pressures by a considerable margin. The persistence of this fourfold structure across all three fluid models, including argon with its experimentally validated non-mean-field critical behavior, confirms that the geometric characterization of the supercritical crossover is robustly sensitive to both the choice of metric and the choice of invariant, and that this sensitivity is a feature of thermodynamic geometry.\\

\section{Conclusions}

In this work we have carried out a systematic analysis of thermodynamic geometry for three fluid models of increasing physical realism: the van der Waals fluid, the Lennard-Jones fluid, and argon described by a multiparameter equation of state fitted to experimental data. The analysis is based on two metric structures, the $V$-metric and the $N$-metric, and on two geometric invariants for each: the scalar curvature $R$ and the scalar curvature density $\mathcal{R} = \sqrt{|g|}\,R$, the latter introduced here as a complementary invariant in the sense of differential geometry. The systematic comparison of these four quantities across three fluid models allows us to distinguish
results that are model-specific from those that reflect genuine features of fluid thermodynamics.\\

In our perspective, the most significant theoretical finding concerns the critical scaling of $\mathcal{R}$. While $R$ diverges as $R \sim t^{-d\nu}$, governed by the correlation length exponent, $\mathcal{R}$ obeys the distinct universal law $\mathcal{R} \sim t^{-(1+\beta)}$, a result that follows analytically from hyperscaling and the Rushbrooke relation and is therefore independent of the universality class. The factor $\sqrt{|g|}$ effectively reweights the curvature by the local thermodynamic volume element, suppressing the dominant contribution of the correlation volume and enhancing the geometric imprint of phase separation. Consequently, $\mathcal{R}$ provides a direct geometric probe of the order parameter exponent $\beta$, a connection entirely absent when only
$R$ is considered. The numerical results confirm this picture across both universality classes examined: mean-field for the Lennard-Jones fluid and Ising for argon.\\

The comparison between the $V$ and $N$-metric constructions reveals that the two metrics agree most closely at intermediate densities and diverge most strongly in the dilute gas and dense liquid limits, a pattern that persists
across all three fluid models and therefore appears to be an intrinsic feature of ensemble-dependent thermodynamic geometry. The loci of exact metric agreement define two new geometric objects introduced here, the Curvature
Equality Curve (CEC) and the Curvature-Density Equality Curve (CDEC), which identify the thermodynamic states where both metric representations yield an identical geometric description of the fluid. The CDEC consistently encloses
a substantially larger region of the phase diagram than the CEC, confirming that the condition $\mathcal{R}_{_{V}} = \mathcal{R}_{_{N}}$ is less restrictive than $R_{_{V}} = R_{_{N}}$. The topology of both curves is qualitatively similar for the Lennard-Jones and argon cases but differs markedly from van der Waals, indicating greater sensitivity to the realism of the equation of state than to the universality class.\\

The generalized $R$-crossing construction shows that the $N$-metric consistently outperforms the $V$-metric in reconstructing the vapor-liquid coexistence curve away from the critical point, particularly along the gas
branch, a noteworthy result given the comparatively limited attention the $N$-metric has received in the literature. The four Widom lines defined by the minima of each invariant converge at the critical point but separate
progressively in the supercritical region in a characteristic fourfold pattern reproduced across all three fluid models, constituting a robust prediction of thermodynamic geometry that is sensitive to both the choice of metric and the choice of invariant.\\

These results establish $\mathcal{R}$ as a physically and geometrically meaningful complement to $R$ in thermodynamic geometry, and demonstrate that the metric choice has consequences extending well beyond the critical region.
The persistence of these findings across models of increasing realism, including argon with experimentally validated Ising critical exponents, supports their robustness as genuine features of fluid thermodynamics. Natural extensions of
this work include the application of scalar curvature densities to other systems where thermodynamic geometry has proven fruitful, such as black holes or magnetic systems, as well as to multiparameter equations of state for other
real substances, where the present framework could be tested. Another avenue worth exploring is the systematic study of model fluids in which the range and softness of the intermolecular potential can be varied continuously, such as Mie or variable-range potentials, allowing one to trace how the geometry of the CEC and CDEC, the structure of the Widom lines, and the critical scaling of $\mathcal{R}$ evolve as the potential parameters are tuned, potentially
revealing universal trends in the ensemble dependence of thermodynamic geometry.\\

\bibliography{biblio}

\end{document}